\documentclass[]{JHEP3}
\usepackage{epsfig,multicol,bbm}

\title{Logarithmic correction to BH entropy as Noether charge}
\author{R Aros, D E D\'{\i}az and A Montecinos\\Universidad Andr\'es Bello,
Departamento de Ciencias F\'{\i}sicas,
Rep\'ublica 220, Santiago, Chile\\
	E-mail: \email{raros,danilodiaz,alejandramontecinos@unab.cl}}

\abstract{We consider the role of the type-A trace anomaly in static black hole solutions to semiclassical Einstein equations in four dimensions.
Via Wald's Noether charge formalism, we compute the contribution to the entropy coming from the anomaly induced effective action and unveil a logarithmic correction to the Bekenstein-Hawking area law.

The corrected entropy is given by a seemingly \emph{universal} formula
$$S_{bh}=\frac{\mathcal{A}_{_H}}{4} - a\cdot\chi_{_H}\cdot \phi_{_H}$$
involving the coefficient $a$ of the \emph{type-A trace anomaly}, the \emph{Euler characteristic} $\chi_{_H}$ of the horizon and the value at the horizon $\phi_{_H}$ of the solution to the \emph{uniformization} problem for \emph{Q-curvature}.
Two instances are examined in detail: Schwarzschild and a four-dimensional massless topological black hole. We also find agreement with the logarithmic correction due to one-loop contribution of conformal fields in the Schwarzschild background.}

\keywords{black hole entropy, trace anomaly, Noether charge}

\begin{document}


\section{Introduction}

It has long been realized that black holes in Einstein gravity behave as thermodynamic objects endowed with Hawking temperature $T_{_H}=\kappa/2\pi$ and intrinsic entropy given by the celebrated Bekenstein-Hawking area law $S_{_{BH}}=A_{_{\mathcal{H}}}/4$. Having set all fundamental constants to unity, one should keep in mind that the previous quantities are the result of taking into account quantum mechanical effects\footnote{The intriguing remark
that \textit{classical} general relativity already ``knew about" \textit{quantum} Hawking radiation somehow parallels a more recent situation, namely that  \textit{classical} Einstein-Hilbert action in $AdS_5$ already ``knew about" the \textit{quantum} trace anomaly of the CFT at the boundary. These questions hint at intricate connections between holography, Hawking radiation and trace anomaly.}, such as Hawking radiation, which made possible (or plausible) the interpretation of the classical laws of black hole mechanics as those of black hole thermodynamics.

In more general gravitational theories, as those containing higher curvature terms in the action, the BH area law no longer applies. Remarkably, assuming the Zeroth Law, a modified form of the First Law  was derived by Wald~\cite{Wald:1993nt} for a wide class of generally covariant actions. Here the role of the entropy is played by the integral of a geometric density, the Noether potential, over a spatial cross-section of the horizon. There are several other ways to compute the macroscopic entropy of a black hole, for instance \cite{Banados:1994qp} uses the connection between angular deficit at the horizon and temperature in the Euclidean context, and it can be proved that many of them are fundamentally equivalent \cite{Iyer:1995kg,Nelson:1994na}.

As counterpart to these macroscopic or thermodynamic approaches, in recent years an increasing number of microscopic or statistical mechanical derivations of black hole entropy has appeared (see, e.g. \cite{Carlip:2008wv} for a review). Of particular interest are those stemming from string theory~\cite{Strominger:1996sh} and quantum geometry~\cite{Ashtekar:1997yu}. The fact that so many different microscopic models yield the same thermodynamic properties has also motivated the search for some kind of symmetry or some other mechanism to explain this ``universality'' (\textit{cf}.~\cite{Fursaev:2004qz}). Carlip~\cite{Carlip:1998wz,Carlip:2008rk} has given an argument, which lies halfway to a microscopic derivation, based on an effective two-dimensional conformal symmetry at the horizon with an associated Virasoro algebra, so that Cardy formula gives a state counting asymptotics in agreement with the BH area law for large central charges. In addition, one can readily find the leading correction to the black hole entropy from Cardy formula, resulting in the \textit{logarithm} of the horizon area~\cite{Carlip:2000nv}.

To come to the subject that we want to examine in this note, further note that the \textit{leading correction} to the entropy in the semiclassical limit of `large' black holes seems to be \textit{universally} given by the logarithm of the horizon area, ( see \textit{e.g.} \cite{Medved:2004eh,Page:2004xp}),
 \begin{equation}
\label{quantumcorrectedEntropy}
 S_{bh} = S_{_{BH}}+ const\cdot\ln S_{_{BH}}+\mathcal{O}(1).
\end{equation}
The different scenarios include one-loop effects of quantum fields near a black hole~\cite{Fursaev:1994te,Mann:1997hm}, quantum gravity/geometry~\cite{Kaul:2000kf}, Carlip's approach already mentioned, partition function of the Euclidean BTZ black hole~\cite{Govindarajan:2001ee} and many others (for a thorough account, we refer to \cite{Medved:2004eh,Page:2004xp}). In a semiclassical framework, where quantum corrections to BH entropy are expected, there is evidence that logarithmic corrections are related to the \textit{trace anomaly}; this happens in the Euclidean formalism after renormalization of one-loop UV-divergent contribution of matter fields to the entropy ~\cite{Fursaev:1994te,Mann:1997hm} and, more recently, in the thermodynamics of certain black hole solutions stemming from an anomalous energy-momentum tensor~\cite{Cai:2009ua}.

Now, perhaps the most notable absence in the above list is a macroscopic derivation based on an effective action. It seems that the leading logarithmic correction to the entropy has eluded Wald's formula. On one hand, the difficulty can be understood due to the lack of a full effective action to account for one-loop effects of quantum fields (let alone higher loops). On the other hand, the higher order (local) curvature invariants pieces of this action only contribute with inverse powers of the area. Yet, the general belief that irrespective of an underlying quantum gravity theory, the limit of sufficiently weak fields at sufficiently long distances should be described by an effective action, makes desirable to have a macroscopic derivation of the leading correction to black hole entropy via Wald's formula.

The purpose of the present note is to use Wald's formula to make a connection between the trace anomaly from quantum one-loop effects of conformal matter fields and a logarithmic correction to the black hole entropy. The rough idea is that the universality of the trace anomaly, which is independent of renormalization schemes and quantum states, should be reflected in a quantum correction to the BH area law. Since the trace of the energy-momentum tensor, the trace anomaly, can be obtained from an ``anomaly-induced effective action'' which can be rendered local via introduction of auxiliary fields, using Wald's formula we can read off its contribution to the entropy as Noether charge. For the cases we consider, we always wind up with a logarithmic term.

The organization of the paper is as follows. We begin in Sec.\ref{SemiclassicalSec} with the semiclassical Einstein equations and focus on the role of the trace anomaly. We then write down an effective action that correctly reproduces the trace part of the Einstein equation. In Sec.\ref{NoetherSec} we proceed to compute the contribution of this anomaly induced effective action to the entropy of black holes solutions of the theory via Wald's Noether charge formalism. Sec.\ref{SchwarzschildSec}  examines the Schwarzschild background and then compares with the result from one-loop renormalization. Sec.\ref{MasslessSec} is devoted to computations in a massless topological black hole. Here we can also solve exactly for the auxiliary field. In Sec.\ref{backreactedSec} we take the backreaction into account and compute the entropy via the First Law to compare with the Noether charge entropy. Finally, in Sec.\ref{DiscussionSec} we summarize and discuss our results and mention some perspectives.

\section{Semiclassical Einstein equations and trace anomaly}\label{SemiclassicalSec}

Our search for the quantum correction to black hole entropy will take place within the framework of quantum field theory in curved backgrounds~\cite{Birrell:1982ix}. The backreaction of quantum fields on the classical spacetime geometry is captured by the semiclassical Einstein equation

\begin{equation}\label{CorrectedEE}
   G_{\mu\nu} + \Lambda g_{\mu\nu}= 8\pi \left\langle T_{\mu\nu} \right\rangle,
\end{equation}
where quantum fields, via the renormalized expectation value of the energy-momentum tensor, `tell spacetime how to curve'.
In four dimensions, contrary to the two-dimensional case, it is a nontrivial task to compute the source term in the semiclassical Einstein equation. Among the many efforts devoted to this endeavor, the most relevant to our present purpose are those that try to approximate the energy-momentum tensor by the ``anomaly induced'' one~\cite{Balbinot:1999vg,Mottola:2006ew}, or equivalently, approximate the would-be one-loop effective action by one whose variation correctly reproduces the anomalous trace of the energy momentum tensor
\begin{equation}\label{traceOfT}
\left\langle T \right\rangle\,\equiv\,-\frac{2}{\sqrt{-g}}\,g^{\mu\nu}\;\frac{\delta\,S_{eff}}{\delta\,g^{\mu\nu}}
\,=\,-\frac{2}{\sqrt{-g}}\,g^{\mu\nu}\;\frac{\delta\,S_{anom}}{\delta\,g^{\mu\nu}} ~.
\end{equation}
The knowledge of the trace and the covariant conservation of the energy-momentum tensor still leaves much freedom associated with the choice of quantum state. This can be thought of as the freedom to add a Weyl-invariant piece to the effective action.

However, for one-loop effects of conformal matter, the trace anomaly is
purely quantum mechanical and it has a \emph{universal} feature, namely, it is independent of the state of the quantum matter fields. Our proposal is then that its effect on the black-hole entropy should equally be captured universally by the anomaly induced effective action upon application of Wald's technique. We follow this intuition and therefore focus on the anomaly induced action.
\subsection*{Anomaly induced action}

In four dimensions, the geometric contribution to the trace anomaly~\cite{DS93} is given by the Euler density (type-A), the Weyl tensor squared (type-B) and a \emph{trivial} part ($\Box R$) that comes from the conformal variation of a local curvature invariant ($R^2$). The type-A and type-B anomalies do not follow from a Lagrangian local in the curvature. The $R^2-$term associated to the trivial anomaly does not produce any logarithm of the area upon application of Wald's technique, therefore we do not pay attention to it. In addition, for the sake of simplicity, we restrict ourselves to the type-A anomaly and consider only the trace given by
\begin{equation}
\left\langle T \right\rangle\,= -\frac{a}{16\pi^2}\,E_4~,
\end{equation}
with $E_{4}=Riem^2-4Ric^2+R^2$.

The standard way to get an action for the trace anomaly is to find a conformal primitive following Polyakov's original computation in two dimensions~\cite{Pol81}. In four dimensions, two generalizations are needed: the conformal properties of the Gaussian curvature and D'Alembertian are inherited by the so-called $Q-$curvature~\cite{BØ91} and the Paneitz's operator~\cite{Paneitz-2008-4}, respectively.
Our preferred $Q-$curvature\footnote{The original Q-curvature \cite{BØ91} differs by a $Weyl^2-$term from the one used here, which is also the one in~\cite{Riegert:1984kt}, but both have the same linear transformation law under Weyl rescaling of the metric.} is given by the combination
\begin{equation}
E_{4} - \frac{2}{3} \Box R~,
\end{equation}
and the Paneitz's operator is
\begin{equation}
\Delta_4=\Box^{\,2}+2\,\nabla_\mu \left( R^{\mu\nu}-\frac{1}{3}g ^{\mu\nu}R\right)\nabla_\nu~.
\end{equation}
This results in a linear transformation law for the Q-curvature under Weyl rescaling of the metric that can be readily integrated to produce a generalized Polyakov's action which, as expected, is nonlocal. In the physics literature, such construction first appeared in~\cite{Riegert:1984kt}, while in conformal geometry the subject has been further developed~\cite{GJMS92,Bra93}
In analogy with Liouville's local form of Polyakov's action, one can introduce an auxiliary field to render the type-A anomaly induced action local, up to boundary terms, (see, \textit{e.g.} \cite{Balbinot:1999vg,Mottola:2006ew})
\begin{eqnarray}\label{EffectiveAction}
^AS_{anom} &=& -\frac{a}{32\pi^2}\int dx^4 \sqrt{-g}\hspace{2ex} \{-\phi\,\Delta_4\,\phi \; +\; Q\,\phi\} \\
&=&-\frac{a}{32\pi^2}\int dx^4 \sqrt{-g}\hspace{2ex} \{-(\Box \phi)^2+2(R^{\mu\nu}-\frac{1}{3}R g^{\mu\nu})\nabla_{\mu}\phi \nabla_{\nu}\phi + Q\phi\} \nonumber.
\end{eqnarray}

The metric variation of this action produces an energy-momentum tensor for the field $\phi$ (the explicit expression can be found in~\cite{Balbinot:1999vg,Mottola:2006ew}). Once the equation of motion for $\phi$ is satisfied
\begin{equation}\label{EffectiveEquationOfPhi}
    \Delta_4 \phi = \frac{1}{2} \left(E_{4}-\frac{2}{3} \Box R \right)~,
\end{equation}
the trace of this anomaly induced energy-momentum tensor equals $E_{4} - \frac{2}{3} \Box R$.
Homogeneous solutions to the equation of motion for $\phi$ only change the traceless part of the semiclassical Einstein equation, and correspond to the specification of the matter quantum state, whereas the trace remains unaffected.
\section{Noether charge of the anomaly induced action}\label{NoetherSec}

In this section we uncover the correction to the usual area law due to
presence of the quantum correction Eq.(\ref{EffectiveAction}). We follow the approach originally proposed by Wald in \cite{Wald:1993nt} where entropy can be computed in terms of a particular Noether charge. In Wald's formulation the manifold has two boundaries, an asymptotic region and the horizon. In these two boundaries one can in principle compute charges; the Noether charges at the asymptotical region corresponds to the mass and angular momenta, etc, whereas the charge at the horizon determines the entropy.
 Previous to the computation of the entropy, one has to deal with the problem of boundary conditions for the possible boundary terms. A proper discussion of this cannot be skipped for the Noether charges are modified by the introduction of such boundary terms. Since our discussion concerns the horizon, for the purposes of this note we assume that the asymptotic charges are properly regularized, for instance \`{a} la~\cite{Olea:2006vd,Aros:2001gz}.

Concerning the horizon, our first step is to fix the boundary term. This is made by declaring equation (\ref{EffectiveAction}) as the action with no further boundary terms at the horizon. With this, we have restricted the family of boundary conditions to a small set, which fortunately is suitable for computing the entropy  with the EH action \cite{Wald:1993nt}.
The entropy in this context can be obtained in terms of a normalized Killing vector, $\hat{\xi}$, that defines the horizon ($\mathcal{H}$) as the surface where
\[
\left.\hat{\xi}^{\mu} \hat{\xi}_{\mu} \right|_{\mathcal{H}} = 0.
\]
$\hat{\xi}$ is normalized such that
\begin{equation}\label{NormalizedKillingVector}
\left.\hat{\xi}^{\mu} \nabla_{\mu} \hat{\xi}^{\nu}\right|_{\mathcal{H}} =2\pi \left.\hat{\xi}^{\nu}\right|_{\mathcal{H}}.
\end{equation}
Next, and following \cite{Wald:1993nt,Jacobson:1993xs}, it can be proved that the variation of the entropy is given by
\[
\delta S = \delta (^{EH} Q(\hat{\xi})) + \delta (^A Q(\hat{\xi})),
\]
where $^{EH} Q(\hat{\xi})$ is the well known Noether charge associated with Einstein Hilbert action \cite{Wald:1993nt}
\[
^{EH} Q(\hat{\xi}) = \frac{1}{16\pi} \int_{\mathcal{H}} \left( \nabla^{\mu} \hat{\xi}^{\nu} \right)  \epsilon_{\mu\nu\alpha\beta} \sqrt{g} dx^{\alpha} \wedge dx^{\beta}
\]
and $^A Q(\hat{\xi})$, the charge associated with the action (\ref{EffectiveAction}), can be split as $^A Q(\hat{\xi}) = Q_{1} + Q_{2} + Q_{3}$ where
\[
Q_{1} = -\frac{a}{16\pi^2} \int_{\mathcal{H}} \left(\nabla^{\lambda} \hat{\xi}^{\mu} \nabla^{\nu} \phi \nabla_{\lambda}\phi -\frac{1}{3} \nabla^{\mu} \hat{\xi}^{\nu} (\nabla \phi)^2 \right)  \epsilon_{\mu\nu\alpha\beta} \sqrt{g} dx^{\alpha} \wedge dx^{\beta},
\]
\[
Q_{2} = \frac{a}{48\pi^2} \int_{\mathcal{H}} \Box \phi \left(\nabla^{\mu} \hat{\xi}^{\nu}  \right) \epsilon_{\mu\nu\alpha\beta} \sqrt{g} dx^{\alpha} \wedge dx^{\beta}
\]
and
\begin{equation}\label{Contributing}
Q_{3} = -\frac{a}{32\pi^2} \int_{\mathcal{H}} \phi \left(\nabla^{\mu} \hat{\xi}^{\nu} R^{\lambda \rho}_{\hspace{2ex} \alpha\beta} \right) \epsilon_{\mu\nu\lambda\rho} \sqrt{g}  dx^{\alpha} \wedge dx^{\beta}.
\end{equation}
Here $\epsilon_{\mu\nu\alpha\beta}$ stands for the Levi-Civita symbol.

As usual, $^{EH} Q(\hat{\xi})$ renders the area law so that the corrections are given by the contributions of $Q_{1}$, $Q_{2}$ and $Q_{3}$. For concreteness, we consider a static geometry, but since our real concern is only the near horizon region this is broad enough to foresee a general behavior that may include stationary geometries as well. This space can be described in Schwarzschild-like coordinates by
\begin{equation}\label{MetricStatic}
ds^2=-f(r)dt^2+\frac{dr^2}{f(r)}+r^2d\Sigma_{\gamma}^2~,
\end{equation}
where the largest zero of $f(r)$ defines the horizon radius $r_{+}$. $\Sigma_{\gamma}$ is the transverse section, which is a compact $\gamma$-constant curvature manifold. We assume it is described by a two dimensional set of coordinates $y^{i}$ and an intrinsic metric $h_{ij}$. In this coordinates it is direct to check that
\begin{equation}\label{Xidef}
 \hat{\xi} = \frac{2\pi}{f(r_{+})'} \partial_{t}.
\end{equation}
It is then easy to see that $Q_1$ vanishes at the horizon and
\[
Q_2 \sim \frac{A_{_\mathcal{H}} }{\beta r_{+}} \left.\frac{d\phi}{dr}\right|_{r_+}
\]
where $\beta^{-1}$ is the temperature of the black hole.
Now we turn our analysis to $Q_{3}$, which reduces to
\[
Q_{3} = -\frac{a}{4\pi} \int_{\mathcal{H}} \phi(r_{+}) \tilde{R}^{ij}_{\hspace{2ex} ij}(y) \sqrt{h} d^2y = -a \cdot\chi_{_\mathcal{H}}\cdot \phi_{_\mathcal{H}}
\]
where $\tilde{R}^{ij}_{\hspace{2ex} kl}$ is the intrinsical two dimensional Riemann tensor and $\chi_{\mathcal{H}}$ is the (2 dimensional) Euler characteristic of the horizon; $\phi_{_\mathcal{H}} = \phi(r_{+})$. The vanishing contribution to the Noether charge from $(\Box\phi)^2$ and the values of $Q_3$ can also be obtained from the cases worked out in~\cite{Jacobson:1993vj}.
In all, the quantum entropy as Noether charge is given by
\begin{equation}\label{ADM-prime}
S_{bh}=\frac{\mathcal{A}_{_\mathcal{H}}}{4} - a\cdot\chi_{_\mathcal{H}}\cdot \phi_{_\mathcal{H}} + \ldots~,
\end{equation}
where the ellipsis stands for terms involving only derivatives of $\phi$.
In the next sections it will be apparent that derivatives of $\phi$, and its powers, translate into inverse powers of the area, so that for the purposes of this note are not relevant. Most remarkable, $\phi_{_\mathcal{H}}$ in $Q_3$ yields a logarithmic correction, $\log(\mathcal{A}_{_\mathcal{H}})$, to the quantum entropy of the black holes.

The boundary conditions for $\phi$ require a little discussion. In principle to solve its equation of motion (\ref{EffectiveEquationOfPhi}), due to its four order nature, one needs to fix four linear independent combinations of $\phi$ and its first three order derivatives on a surface to have a proper set of boundary conditions. Another option is two set of two linear independent relations on two non intersecting surfaces. In the case at hand, which is static and with radial symmetry, this corresponds to fix four relations on a $r=$const surface or two conditions on two $r$=const. surfaces. This is very helpful to understand why, even though it is not enough to determine $\phi$ globally, since our concern is only the near horizon region, we will only require the finiteness of $\phi$ and its first derivative at the horizon.
\section{Schwarzschild black hole}\label{SchwarzschildSec}

For obvious (or historical) reasons, we first work out the proposed entropy correction for Schwarzschild black hole
\begin{equation}\label{Schwarzschild}
ds^2=-\left(1-\frac{2m}{r}\right)dt^2+\left(1-\frac{2m}{r}\right)^{-1} dr^2+r^2d\Omega^2~.
\end{equation}
We only have to solve the equation of motion (\ref{EffectiveEquationOfPhi}) for $\phi$, which greatly simplifies in this Ricci-flat background
\begin{equation}\label{Unif-Schwarzschild}
    \Box^{\,2} \phi = \frac{24\,m^2}{r^6}~.
\end{equation}
In order to apply Wald's formula, we require regularity of $\phi$ and its derivative at the horizon and look for a static spherically symmetric solution. It is remarkable that an explicit solution satisfying these requirements can be found~\cite{Balbinot:1999vg,Mottola:2006ew}
\begin{equation}
\frac{d\phi}{dr}= -\frac{4m}{3r(r-2m)}\ln\frac{r}{2m}-\frac{1}{2m}-\frac{2}{r}+C_{\infty}\left(r+2m+\frac{4m^2}{r}\right)~.
\end{equation}
The integration constant $C_{\infty}$ is what remains of the solution to the homogeneous equation and it is tuned to zero. Two apparent reasons for this choice are the following. First, the term linear in $r$ becomes $1/\widehat{r}$ after inversion\footnote{Notice that the equation for $\phi$ transforms `nicely' under the conformal group and, as Schwarzschild is asymptotically flat, also under inversion.} and introduces a spurious source at $\widehat{r}=0$. Second, the energy-momentum tensor only contains derivatives of $\phi$ so that a constant term at spatial infinity is `physically' admissible (thermal bath) but not one diverging with $r$. This very same choice $C_{\infty}=0$ is considered in \cite{Mottola:2006ew}, but differs from the one in \cite{Balbinot:1999vg}.
Upon integration we get
\begin{equation}
\phi(r)=-\frac{r}{2m}-2\ln r+\frac{1}{3}\left(\ln\frac{r}{2m}\right)^2+\frac{2}{3}\,\textrm{dilog} \left(\frac{r}{2m}\right)~.
\end{equation}
Any constant term would change the value at the horizon of $\phi$ but its contribution to the Noether charge would then vanish because the action reduces to a topological term, the integral of Euler density. Therefore we only keep the $r$-dependent part, and any constant term in the final expression is absorbed in an $\mathcal{O}(1)$ term. The crucial terms is then $-2\ln r$, that at the horizon gives $-\ln A_{_H}$. The contribution of the anomaly induced effective action to the Noether potential, equation (\ref{ADM-prime}) with $\chi_{_H}=2$ for a spherical horizon, results then in a logarithmic correction to the black hole entropy
\begin{equation}
S_{bh}=S_{BH} + 2\,a\cdot\ln S_{BH} + \mathcal{O}(1)~.
\end{equation}

\subsection*{Comparison with quantum one-loop correction}

The anomalous trace of the energy-momentum tensor from one-loop effects of (conformal) quantum matter comes from the trace anomaly of the functional determinants of the inverse propagators. It depends on the number $N_s$ of massless fields of spin $s$~(\textit{cf}. \cite{Birrell:1982ix,Christensen:1978gi}). For spins $(0,1/2,1)$ corresponding to real scalars, Dirac spinors and gauge fields, respectively, one has
\begin{equation}
\left\langle T \right\rangle\,= \frac{1}{16\pi^2}\,\left\{c\, C^2-a\, E_4\right\}~,
\end{equation}
with $C$ being the Weyl tensor and
\begin{eqnarray}
a&=&\frac{1}{360}(N_0+11N_{1/2}+62N_1)~,\\\nonumber\\
c&=&\frac{1}{120}(N_0+6N_{1/2}+12N_1)~.
\end{eqnarray}
The correction to the entropy after renormalization in the Euclidean formulation as computed in~\cite{Fursaev:1994te} is given by\footnote{The apparent discrepancy with Fursaev's result is simply due to the fact that his $N_{1/2}$ counts Weyl or Majorana spinors.}
\begin{equation}
S_{bh}=S_{BH}+\frac{2N_0+7N_{1/2}-26N_1}{180}\cdot\ln S_{BH} + \mathcal{O}(1)~.
\end{equation}
This correction agrees with the result reported in~\cite{Solodukhin:1997yy}  based on dimensional and scaling arguments
\begin{equation}\label{Solodukhin}
S_{bh}=S_{BH}-\frac{b}{2}\cdot\ln S_{BH} + \mathcal{O}(1)~,
\end{equation}
with
\begin{equation}
b=\int d^4x\sqrt{g}\left\langle T \right\rangle
\end{equation}
where the integral is taken over the Euclidean black hole, whose Euler number is 2.

Although we only considered type-A anomaly, our result can be adapted to the above case because for Schwarzschild and any other Ricci-flat background both terms $E_4$ and $C^2$ coincide. We only have to make a shift in our entry $a\rightarrow (a-c)$ to absorb the $C^2-$term in the $E_4-$tem. We find then perfect agreement.

\section{Massless topological black hole}\label{MasslessSec}

The second example to consider is the four dimensional massless case of topological black holes (\textit{cf.} \cite{Aros:2000ij})
\begin{equation}\label{TBH}
ds^2=-\left(\frac{r^2}{r_{_+}^2}-1\right)dt^2+ \left(\frac{r^2}{r_{_+}^2}-1\right)^{-1} dr^2 + r^2d\Sigma_-^2,
\end{equation}
where $d\Sigma_-^2$ is the line element of a two-dimensional compact surface of negative constant curvature.
The motivation being that the high degree of symmetry of this background, locally $AdS$, could simplify the calculations.
Being locally AdS, $R^{\alpha\beta}_{\hspace{2ex} \mu \nu} = r_{+}^{-2} \delta^{\alpha\beta}_{\mu \nu}$, the Weyl tensor vanishes and therefore only type-A trace anomaly is to be considered, $E_4=24 r_{+}^{-4}$.
The equation of motion (\ref{EffectiveEquationOfPhi}) for $\phi$ reduces to
\begin{equation}
\left(\Box^{\,2} + \frac{2}{r_{_+}^{2}}\, \Box \right) \phi = \frac{12}{r_{_+}^{4}}.
\end{equation}
Imposing regularity of $\phi$  and its derivative at the horizon and considering only radial dependence, the solution reads
\begin{equation}
\phi (r)= 2 \ln(r + r_{_+})+ C_1 + \frac{(-2 r_{_+}+ C_2)}{r}~.
\end{equation}
Here $C_1$ and $C_2$ are remaining constants from the homogeneous solution, two other integration constants were already used to meet the regularity conditions at the horizon. $C_2$ is tuned to avoid a delta-like artifact at $r=0$ and $C_1$ is not relevant, as in the Schwarzschild case. Therefore,
\begin{equation}
\phi(r_{_+})=2\,\ln r_{_+}+\mathcal{O}(1)=\ln A_{_H} + \mathcal{O}(1).
\end{equation}
Finally, back to the formula (\ref{ADM-prime}), the corrected entropy  is given by
\begin{equation}
    S_{bh} = S_{_{BH}} -a\cdot\chi_{_H}\cdot\ln S_{_{BH}}+\mathcal{O}(1)~,
\end{equation}
where now $\chi_{_H}=2-2g<0$ is the (negative) Euler number of the horizon.
Notice that in this case, the formula (\ref{Solodukhin}) cannot be applied, at least straightforwardly; since $E_4$ is constant, one faces the infinite volume of the Euclidean black hole and some regularization must be done.

\section{Backreaction and First Law}\label{backreactedSec}

One further step in the semiclassical program would be to plug the backreacted metric into the equation of motion for $\phi$. In general, no explicit solution for the metric is at hand for loop-effect of quantum fields and usually some kind of approximation is involved. However, quite recently, a static black hole solution has been found for the Einstein equation with an anomalous energy-momentum tensor~\cite{Cai:2009ua}. The solution to the trace of Einstein equation gives a $g_{tt}=-1/g_{rr}$ while the traceless components of the energy-momentum tensor are chosen somewhat \textit{ad hoc} to support the static ansatz. The First Law applied to the solution gives a logarithmic correction to the black hole entropy.

Despite some misgivings, we follow this approach and consider, in addition, a cosmological constant $\Lambda$ and generic compact constant curvature horizon with curvature $2\gamma=-2,0,2$. We consider the metric ansatz
\begin{equation}\label{GenCCO}
ds^2=-f(r)dt^2+\frac{dr^2}{f(r)}+r^2d\Sigma_{\gamma}^2~,
\end{equation}
to look for solutions to the trace of the semiclassical Einstein equation
\begin{equation}\label{Trace-Backreact}
-R+4\Lambda=8\pi\left\langle T \right\rangle~,
\end{equation}
with trace anomaly
\begin{equation}
\left\langle T \right\rangle\,= -\frac{a}{16\pi^2}\,E_4~.
\end{equation}
The explicit solution reads
\begin{equation}\label{improvedBackreacted}
f(r)=\gamma -{\frac {3\,{r}^{2} - \sqrt {9\,{r}^{4}+
144 \alpha^{2}{{\it \gamma}}^{2}-24\, \alpha\,\Lambda\,{r}^{4}+72\,
\alpha \, C_2-72\,\alpha\,C_1\,r}}{12 \alpha }}~,
\end{equation}
where $\alpha\equiv a/2\pi$.
This solution is a generalization of the solution studied in \cite{Cai:2009ua}. With $C_2=-2\alpha\gamma$ and $C_1=2\mu$, the limit $a\rightarrow 0$ reduces to several known geometries depending on the values of $(\gamma,\Lambda)$ being larger, equal or less than zero: Schwarzschild ($>,=$), topological black hole ($<,<$), Schwarzschild-AdS ($>,<$) and Schwarzschild-dS ($>,>$).

The entropy of the black hole (\ref{improvedBackreacted}) can be deduced from the first law of thermodynamics. The mass of this black hole $M\equiv\mu\frac{\Sigma}{4\pi}$, with $\Sigma$ the area of the transverse section (hyperbolic, spherical or toroidal), can be expressed as function of the radius of the horizon $r_{+}$ as
\begin{equation}
M=\frac{\Sigma}{4\pi}\,\frac{r_+}{2}\left( \gamma-\frac{2\gamma^2 \alpha}{r_+^2}-\frac{\Lambda r_+^2}{3}\right).
\end{equation}
Analogously, the temperature is given by
\begin{equation}
T=\frac{1}{4\pi(r_+^2-4\gamma \alpha)}\left(\gamma r_+ -\Lambda r_+^3+\frac{2 \alpha\gamma ^2}{r_+}\right)
\end{equation}
The first law of thermodynamics in this case allows to write the entropy as the integral along $r_+$
\begin{equation}
S=\int\frac{1}{T}\frac{dM}{dr_+}dr_+.
\end{equation}
This yields
\begin{equation}
S_{bh}=\frac{\Sigma\; r_+^2}{4}\,-\Sigma\gamma \alpha\ln r_+^2
\end{equation}
modulo an integration constant, so that
\begin{equation}
S_{bh}=S_{BH}-a\cdot \chi_{_H}\cdot \ln S_{BH}+\mathcal{O}(1)~.
\end{equation}

At this point, to compare with our entropy correction we should solve the equation (\ref{EffectiveEquationOfPhi}) for $\phi$ using (\ref{improvedBackreacted}).
For the massless topological we are able to solve the equation for the auxiliary field because backreaction on the geometry (\ref{Trace-Backreact}) can be absorbed in the cosmological constant. We have not been able to solve in the other cases yet, but we observe an intriguing feature: backreaction in the massless topological black hole does not alter the log-term whereas for Schwarzschild there is apparently a non-analyticity in the limit $a\rightarrow 0$, a sign change.
\section{Discussion}\label{DiscussionSec}

By computing the Noether charge of the anomaly induced action, we have obtained a correction to black hole entropy directly connected with the type-A trace anomaly. The correction has a purely geometric content, apparent from the factors involving the Euler characteristic of the horizon and the coefficient of the type-A geometric contribution to the trace anomaly. But the auxiliary field $\phi$ has also an interesting interpretation in conformal geometry, namely, it corresponds to the solution to the \textit{uniformization} problem for the Q-curvature (see, \textit{e.g.} \cite{malchiodi-2007-3} for compact Riemannian manifolds) in the particular case where the constant values of the Weyl-transformed Q-curvature is zero.
We have only been able to compute $\phi$ for Schwarzschild and the massless topological black hole; in general, we expect that the log-term should be somehow recovered form the near-horizon region and yet, the value of $\phi$ still has a flavor to non-locality.

In relation with our findings, several issues deserve further examination: appropriate treatment of type-B anomaly, possible generalization to higher even dimensions, the curious analogy with black hole solutions in Horava-Lifshitz gravity~\cite{Cai:2009pe}, uniformization problem in Lorentzian signature, dimensional reduction to (1+1)-dimensions, analogy with two-dimensional dilaton-gravity, careful analysis of boundary terms, and the case of extremal black holes.

The logarithmic correction is by far not the whole story, but our result (if correct!) would be a consistency check for any derivation involving the trace anomaly, very much in the spirit of the  Bekenstein-Hawking area law.

\section*{Acknowledgments}

This work was partially funded through Fondecyt-Chile 3090012, UNAB AR-01-09 and UNAB DI 2009/04.
We would like to express our deep sympathy to all those affected by the recent tragic events in Chile.


\hspace{0.5cm}

\begin{thebibliography}{10}

\bibitem{Wald:1993nt}
R.~M. Wald, {\it Black Hole Entropy is Noether Charge},  {\em Phys. Rev.} {\bf
  D48} (1993) 3427--3431, [\href{http://xxx.lanl.gov/abs/gr-qc/9307038}{{\tt
  gr-qc/9307038}}].

\bibitem{Banados:1994qp}
M.~Banados, C.~Teitelboim, and J.~Zanelli, {\it Black Hole Entropy and the Dimensional Continuation of the Gauss-Bonnet Theorem},  {\em Phys. Rev.
  Lett.} {\bf 72} (1994) 957--960,
  [\href{http://xxx.lanl.gov/abs/gr-qc/9309026}{{\tt
  gr-qc/9309026}}].

\bibitem{Iyer:1995kg}
V.~Iyer and R.~M. Wald, {\it A comparison of noether charge and euclidean
  methods for computing the entropy of stationary black holes},  {\em Phys.
  Rev.} {\bf D52} (1995) 4430--4439,
  [\href{http://xxx.lanl.gov/abs/gr-qc/9503052}{{\tt gr-qc/9503052}}].

\bibitem{Nelson:1994na}
W.~Nelson, {\it {A Comment on black hole entropy in string theory}},  {\em
  Phys. Rev.} {\bf D50} (1994) 7400--7402,
  [\href{http://xxx.lanl.gov/abs/hep-th/9406011}{{\tt hep-th/9406011}}].

\bibitem{Carlip:2008wv}
  S.~Carlip,
  ``Black Hole Thermodynamics and Statistical Mechanics,''
  Lect.\ Notes Phys.\  {\bf 769} (2009) 89
  [arXiv:0807.4520 [gr-qc]].

\bibitem{Strominger:1996sh}
  A.~Strominger and C.~Vafa,
  {\it Microscopic Origin of the Bekenstein-Hawking Entropy,}
 {\em Phys.\ Lett.\  B} {\bf 379} (1996) 99
  [arXiv:hep-th/9601029].


\bibitem{Ashtekar:1997yu}
  A.~Ashtekar, J.~Baez, A.~Corichi and K.~Krasnov,
  {\it Quantum geometry and black hole entropy,}
  {\em Phys.\ Rev.\ Lett.}  {\bf 80} (1998) 904
  [arXiv:gr-qc/9710007].

\bibitem{Fursaev:2004qz}
  D.~V.~Fursaev,
  {\it Can one understand black hole entropy without knowing much about quantum
  gravity?,} {\em  Phys.\ Part.\ Nucl.}  {\bf 36}, 81 (2005)
  [{\em Fiz.\ Elem.\ Chast.\ Atom.\ Yadra} {\bf 36}, 146 (2005)]
  [arXiv:gr-qc/0404038].

\bibitem{Carlip:1998wz}
S.~Carlip, {\it Black hole entropy from conformal field theory in any
  dimension},  {\em Phys. Rev. Lett.} {\bf 82} (1999) 2828--2831,
  [\href{http://xxx.lanl.gov/abs/hep-th/9812013}{{\tt
  hep-th/9812013}}].

\bibitem{Carlip:2008rk}
  S.~Carlip,
  ``Black Hole Entropy and the Problem of Universality,''
  arXiv:0807.4192 [gr-qc].


\bibitem{Carlip:2000nv}
  S.~Carlip,
  ``Logarithmic corrections to black hole entropy from the Cardy formula,''
  Class.\ Quant.\ Grav.\  {\bf 17} (2000) 4175
  [arXiv:gr-qc/0005017].

\bibitem{Medved:2004eh}
A.~J.~M. Medved, {\it {A comment on black hole entropy or why Nature abhors a
  logarithm}},  {\em Class. Quant. Grav.} {\bf 22} (2005) 133--142,
  [\href{http://xxx.lanl.gov/abs/gr-qc/0406044}{{\tt gr-qc/0406044}}].

\bibitem{Page:2004xp}
  D.~N.~Page,
  {\it Hawking radiation and black hole thermodynamics,}{\em
  New J.\ Phys.}  {\bf 7}, 203 (2005)
  [\href{http://xxx.lanl.gov/abs/hep-th/0409024}{{\tt gr-qc/0409024}}].

\bibitem{Fursaev:1994te}
  D.~V.~Fursaev,
  ``Temperature And Entropy Of A Quantum Black Hole And Conformal Anomaly,''
  Phys.\ Rev.\  D {\bf 51} (1995) 5352
  [arXiv:hep-th/9412161].

\bibitem{Mann:1997hm}
  R.~B.~Mann and S.~N.~Solodukhin,
  {\it Universality of quantum entropy for extreme black holes,}
  {\em Nucl.\ Phys.\  B} {\bf 523}, 293 (1998)
  [arXiv:hep-th/9709064].

\bibitem{Kaul:2000kf}
  R.~K.~Kaul and P.~Majumdar,
{\it Logarithmic correction to the Bekenstein-Hawking entropy,}
{\em Phys.\ Rev.\ Lett.}  {\bf 84}, 5255 (2000)
  [arXiv:gr-qc/0002040].

\bibitem{Govindarajan:2001ee}
  T.~R.~Govindarajan, R.~K.~Kaul and V.~Suneeta,
  {\it Logarithmic correction to the Bekenstein-Hawking entropy of the BTZ  black
  hole,}
  {\em Class.\ Quant.\ Grav.} {\bf 18}, 2877 (2001)
  [arXiv:gr-qc/0104010].

\bibitem{Cai:2009ua}
R.-G. Cai, L.-M. Cao, and N.~Ohta, {\it {Black Holes in Gravity with Conformal
  Anomaly and Logarithmic Term in Black Hole Entropy}},
  \href{http://xxx.lanl.gov/abs/0911.4379}{{\tt 0911.4379}}.

\bibitem{Birrell:1982ix}
  N.~D.~Birrell and P.~C.~W.~Davies,
  {\it Quantum Fields In Curved Space,}
{\em  Cambridge, Uk: Univ. Pr.} ( 1982) 340

\bibitem{Balbinot:1999vg}
R.~Balbinot, A.~Fabbri, and I.~L. Shapiro, {\it {Vacuum polarization in
  Schwarzschild space-time by anomaly induced effective actions}},  {\em Nucl.
  Phys.} {\bf B559} (1999) 301--319,
  [\href{http://xxx.lanl.gov/abs/hep-th/9904162}{{\tt hep-th/9904162}}].

\bibitem{Mottola:2006ew}
E.~Mottola and R.~Vaulin, {\it Macroscopic effects of the quantum trace
  anomaly},  {\em Phys. Rev.} {\bf D74} (2006) 064004,
  [\href{http://xxx.lanl.gov/abs/gr-qc/0604051}{{\tt gr-qc/0604051}}].

\bibitem{DS93}
S.~Deser and A.~Schwimmer, ``Geometric classification of conformal
anomalies in arbitrary dimensions,'' Phys.\ Lett.\  B {\bf 309}
(1993) 279 [arXiv:hep-th/9302047].

\bibitem{Pol81}
A.~M.~Polyakov, ``Quantum geometry of bosonic strings,'' Phys.\
Lett.\  B {\bf 103} (1981) 207.

\bibitem{BØ91}
T. ~Branson and B. ~Ørsted,
``Explicit functional determinants in four dimensions,''
Proc.\ Amer.\ Math.\ Soc.\ {\bf 113} (1991) 669.

\bibitem{Paneitz-2008-4}
S.~M. Paneitz, {\it A quartic conformally covariant differential operator for
  arbitrary pseudo-riemannian manifolds (summary)},  {\em SIGMA} {\bf 4} (2008)
  036.

\bibitem{Riegert:1984kt}
R.~J. Riegert, {\it A nonlocal action for the trace anomaly},  {\em Phys.
  Lett.} {\bf B134} (1984) 56--60.


\bibitem{GJMS92}
C.~R.~Graham, R.~Jenne, L.~J.~Manson, G.~A.~J.~Sparling,
``Conformally invariant powers of the Laplacian. I. Existence,'' J.\
London\ Math.\ Soc. (2) {\bf 46} (1992) 557.

\bibitem{Bra93}
T.~Branson, ``The Functional Determinant,'' Global\ Analysis\
Research\ Center\ Lecture\ Note\ Series, Number 4, Seoul\ National\
University (1993);
``Sharp inequalities, the functional determinant, and
the complementary series,'' Trans.\ Amer.\ Math.\ Soc. {\bf 347}
(1995) 3671.



\bibitem{Olea:2006vd}
R.~Olea, {\it {Regularization of odd-dimensional AdS gravity: Kounterterms}},
  {\em JHEP} {\bf 04} (2007) 073,
  [\href{http://xxx.lanl.gov/abs/hep-th/0610230}{{\tt hep-th/0610230}}].

\bibitem{Jacobson:1993xs}
T.~Jacobson and R.~C. Myers, {\it Black hole entropy and higher curvature
  interactions},  {\em Phys. Rev. Lett.} {\bf 70} (1993) 3684--3687,
  [\href{http://xxx.lanl.gov/abs/hep-th/9305016}{{\tt hep-th/9305016}}].

\bibitem{Aros:2001gz}
R.~Aros, {\it Analyzing charges in even dimensions},  {\em Class. Quant. Grav.}
  {\bf 18} (2001) 5359--5369,
  [\href{http://xxx.lanl.gov/abs/gr-qc/0011009}{{\tt gr-qc/0011009}}].

\bibitem{Jacobson:1993vj}
  T.~Jacobson, G.~Kang and R.~C.~Myers,
  Phys.\ Rev.\  D {\bf 49} (1994) 6587
  [arXiv:gr-qc/9312023].


\bibitem{Christensen:1978gi}
  S.~M.~Christensen and M.~J.~Duff,
  ``Axial And Conformal Anomalies For Arbitrary Spin In Gravity And
  Supergravity,''
  Phys.\ Lett.\  B {\bf 76} (1978) 571.

\bibitem{Solodukhin:1997yy}
  S.~N.~Solodukhin,
  {\it Entropy of Schwarzschild black hole and string-black hole
  correspondence,}
  {\em Phys.\ Rev.\  D} {\bf 57}, 2410 (1998)
  [arXiv:hep-th/9701106].

\bibitem{Aros:2000ij}
R.~Aros, R.~Troncoso, and J.~Zanelli, {\it Black holes with topologically
  nontrivial ads asymptotics},  {\em Phys. Rev.} {\bf D63} (2001) 084015,
  [\href{http://xxx.lanl.gov/abs/hep-th/0011097}{{\tt hep-th/0011097}}].

\bibitem{malchiodi-2007-3}
A. ~Malchiodi,
``Conformal Metrics with Constant Q-Curvature,''
SIGMA 3,(2001)120.


\bibitem{Cai:2009pe}
  R.~G.~Cai, L.~M.~Cao and N.~Ohta,
  ``Topological Black Holes in Horava-Lifshitz Gravity,''
  Phys.\ Rev.\  D {\bf 80} (2009) 024003
  [arXiv:0904.3670 [hep-th]].

\end{thebibliography}

\providecommand{\href}[2]{#2}\begingroup\raggedright\endgroup
\end{document}